\documentclass[prl,aps,twocolumn,showpacs]{revtex4}

\usepackage{amsmath}
\usepackage{bm}
\usepackage{graphicx}

\begin{document}

\title{Split-merge cycle, fragmented collapse, and vortex disintegration
in rotating Bose-Einstein condensates with attractive interactions}

\author{Hiroki Saito}
\author{Masahito Ueda}
\affiliation{Department of Physics, Tokyo Institute of Technology,
Tokyo 152-8551, Japan \\
and CREST, Japan Science and Technology Corporation (JST), Saitama
332-0012, Japan}

\date{\today}

\begin{abstract}
The dynamical instabilities and ensuing dynamics of singly- and
doubly-quantized vortex states of Bose-Einstein condensates with
attractive interactions are investigated using full 3D numerical
simulations of the Gross-Pitaevskii equation.
With increasing the strength of attractive interactions, a series of
dynamical instabilities such as quadrupole, dipole, octupole, and monopole
instabilities emerge.
The most prominent instability depends on the strength of interactions,
the geometry of the trapping potential, and deviations from the
axisymmetry due to external perturbations.
Singly-quantized vortices split into two clusters and subsequently
undergo split-merge cycles in a pancake-shaped trap, whereas the split
fragments immediately collapse in a spherical trap.
Doubly-quantized vortices are always unstable to disintegration of the
vortex core.
If we suddenly change the strength of interaction to within a certain
range, the vortex splits into three clusters, and one of the clusters
collapses after a few split-merge cycles.
The vortex split can be observed using a current experimental setup of the
MIT group.
\end{abstract}


\maketitle

\section{Introduction}

Quantized vortices are among the hallmarks of superfluidity, and have been
observed in the gaseous Bose-Einstein condensates (BECs) of ${}^{87}{\rm
Rb}$~\cite{Matthews,Madison} and ${}^{23}{\rm Na}$~\cite{Abo}, where
repulsive interatomic interactions play a crucial role in nucleation and
stabilization of vortices.
In the case of attractive interactions, however, creating vortices by
stirring the condensate in a harmonic trap is difficult because the vortex
state is energetically unfavorable against decaying into the non-vortex
ground state.
A possible experimental scheme for creating a vortex state with attractive
interactions is to prepare it in the repulsive regime and then switch the
sign of the interaction to attractive by using, e.g., the Feshbach
resonance~\cite{Inouye,Cornish}.
Another possible scheme to create vortices in an attractive BEC is to
imprint topological phases on the condensate using either controlled
interconversion between two components~\cite{Williams,Matthews} or spin
rotation by adiabatic change of the magnetic
field~\cite{Nakahara,Leanhardt}.
The aim of the present paper is to study the dynamical instabilities and
ensuing dynamics of vortex states with attractive interactions in 3D
axisymmetric traps.

There are two types of instabilities that cause a vortex state to decay:
dynamical and thermodynamic instabilities.
Dynamical instabilities, or modulational instabilities, set in when the
Bogoliubov spectrum acquires an imaginary part which causes an exponential
growth in the corresponding mode.
Vortex nucleation~\cite{Recati,Sinha}, vortex exchange between two
components~\cite{Garcia00}, splitting of a vortex with attractive
interactions~\cite{SaitoL}, and disintegration of a doubly-quantized
vortex with repulsive interactions~\cite{Garcia99,Mottonen} are all
attributed to dynamical instabilities.
Thermodynamic instabilities are caused by interactions between the
condensate and the thermal cloud accompanied by dissipation of the energy
and angular momentum.
Reductions in the number of vortices~\cite{Madison}, vortex
bending~\cite{Rosen,Garciabend}, and vortex-lattice formation~\cite{Abo}
are associated with thermodynamic instabilities.
The time scale of the latter instabilities at low temperatures is
typically $\sim 1$ s or longer~\cite{Madison,Abo,Engels}.
In contrast, the time scale of dynamical instabilities can be much faster
than that of thermodynamic instabilities since the former grows
exponentially in time.
We shall therefore focus on dynamical instabilities in this paper.

In Ref.~\cite{Pu}, Pu {\it et al.} reported that singly- and
doubly-quantized vortices in 2D exhibit a series of dynamical
instabilities when the strength of interaction exceeds the corresponding
critical values.
We numerically solved the time-dependent Gross-Pitaevskii (GP) equation
for a 2D system, and showed that the vortex splits into two clusters which
revolve around the center of the trap, and undergo split-merge cycles or
collapse~\cite{SaitoL}.
In the present paper, we examine such dynamics in full 3D systems and
also study the dynamical instability of the doubly-quantized vortex, that
was recently realized by the MIT group~\cite{Leanhardt}.
Our primary findings are the following: \\
\ (1) Our 3D analysis reveals that the dynamical instabilities depend
crucially on the trap geometry.
For a singly-quantized vortex, split-merge cycles can occur in a
pancake-shaped trap, whereas the split fragments immediately collapse in a
spherical trap.
In a cigar-shaped trap, there is a parameter range in which the vortex can
collapse into a single cluster due to the dipole instability. \\
\ (2) The doubly-quantized vortex with attractive interactions is always
unstable against disintegration of the vortex core.
If the strength of interaction is suddenly changed to within a certain
range, the vortex splits into three clusters which undergo split-merge
cycles before one or possibly more of them collapse. \\
\ (3) Even when the lifetime of the condensate is very short, the
phenomenon of vortex split can be observed by suddenly changing the 
interaction to a large attractive value.
The split of a doubly-quantized vortex is shown to be observable using
a current experimental setup of the MIT group. \\
\ (4) Using a variational method, we find that both the split instability
and the split-merge cycles can be understood by using only three basis
functions. \\
\ (5) In the immediate vicinity of the critical strength of interaction
for the vortex split, quantum fluctuations like the ones found in a 1D
case~\cite{Kanamoto} become prominent.

This paper is organized as follows.
Section~\ref{s:formulation} briefly reviews the GP and Bogoliubov theories
and discusses dynamical instabilities.
Section~\ref{s:single} studies dynamical instabilities of a
singly-quantized vortex state, and Sec.~\ref{s:double} those of a
doubly-quantized vortex state.
Section~\ref{s:variational} examines a variational method to understand
the dynamical instabilities and split-merge phenomenon.
Section~\ref{s:lowly} shows that the split-merge phenomenon is accompanied
by the emergence of a symmetry-broken low-lying state.
Section~\ref{s:conclusion} concludes this paper.

\section{Formulation of the problem}
\label{s:formulation}

We consider a dilute-gas BEC at zero temperature confined in an
axisymmetric harmonic trap.
The dynamics of the condensate are described by the GP equation
\begin{equation} \label{GP}
i \hbar \frac{\partial \psi}{\partial t} = -\frac{\hbar^2}{2m} \nabla^2
\psi + V_{\rm trap} \psi + \frac{4\pi \hbar^2 a}{m} |\psi|^2 \psi,
\end{equation}
where $m$ is the atomic mass, $V_{\rm trap} \equiv m \omega_\perp^2 (x^2 +
y^2) / 2 + m \omega_z^2 / 2$ with $\omega_\perp$ and $\omega_z$ being the
radial and axial trap frequencies, and $a$ is the s-wave scattering
length.
We normalize the length, time, energy, and wave function by $d_0 \equiv
(\hbar / m \omega_\perp)^{1/2}$, $\omega_\perp^{-1}$, $\hbar
\omega_\perp$, and $(N / d_0^3)^{1/2}$, respectively, where $N$ is the
number of atoms in the BEC, and the wave function is normalized as $\int
d{\bf r} |\psi|^2 = 1$.
The GP equation (\ref{GP}) is then written as
\begin{equation} \label{GPN}
i \frac{\partial \psi}{\partial t} = -\frac{1}{2} \nabla^2
\psi + \frac{1}{2} (r^2 + \lambda^2 z^2) \psi + g |\psi|^2 \psi,
\end{equation}
where $r^2 \equiv x^2 + y^2$, $\lambda \equiv \omega_z / \omega_\perp$,
and $g \equiv 4\pi N a / d_0$.

The Bogoliubov spectrum for a stationary state $\psi_0$ of the GP
equation is obtained from the Bogoliubov-de Gennes equations
\begin{subequations} \label{Bogo}
\begin{eqnarray}
\left( K + 2 g |\psi_0|^2 \right) u_n + g \psi_0^2 v_n & = & E_n u_n, \\
\left( K + 2 g |\psi_0|^2 \right) v_n + g \psi_0^{*2} u_n & = & -E_n v_n,
\end{eqnarray}
\end{subequations}
where $K \equiv -\nabla^2 / 2 + (r^2 + \lambda^2 z^2) / 2 - \mu$ and $\mu$
is the chemical potential.
For the axisymmetric vortex states $\psi_0 \propto e^{i q \phi}$, each
Bogoliubov mode $u_n \propto e^{i (q + \delta m) \phi}$ couples only to
$v_n \propto e^{-i (q - \delta m) \phi}$, where $q$ and $\delta m$ are
integers and $\phi = \tan^{-1} y / x$ is the azimuthal angle.
In the following discussions, we shall refer to the modes with $\delta m =
1$, $2$, and $3$ as the dipole, quadrupole, and octupole modes,
respectively.

The stability of the system is characterized by the Bogoliubov spectrum
$E_n$.
When the system is in the ground state, or in a metastable state for which
there is an energy barrier that prevents the system from decaying into a
lower energy state, all Bogoliubov eigenenergies are real and positive.
An example of such a metastable state is a BEC with an attractive
interaction below the critical strength of interaction for
collapse~\cite{Sackett97,Ueda98,SaitoA01}.
When there are negative eigenvalues in the Bogoliubov spectrum, the system
can thermodynamically decay into a lower energy state by dissipating 
energy to the environment.
When complex eigenvalues emerge in the Bogoliubov spectrum, the
corresponding modes grow exponentially in time and the system becomes
dynamically unstable.
The time scale of such dynamical instability is given by the inverse of
the imaginary part.
The difference between thermodynamic and dynamical instabilities is
that the former requires energy dissipation, whereas the latter can occur
via an energy-conserving process.
Thus, the dynamical instability can be a dominant decay process at
sufficiently low temperature in a high-vacuum chamber in which dissipation
is negligible over a short time.

The time evolution of a system that deviates slightly from the
stationary state $\psi_0$ of the GP equation can be described in terms of
the Bogoliubov modes $u_n$ and $v_n$ as~\cite{Edwards}
\begin{equation}
\psi(t) = e^{-i \mu t} \left[ \psi_0 + \sum_n (\varepsilon_n u_n e^{-i
E_n t} + \varepsilon_n^* v_n^* e^{i E_n^* t}) \right],
\end{equation}
where $\varepsilon_n$'s ($|\varepsilon_n| \ll 1$) are small coefficients
that describe the deviation.
When the eigenvalue of a Bogoliubov mode becomes complex, the mode grows
exponentially in time as
\begin{widetext}
\begin{eqnarray} \label{modegrow}
\psi(t) & \simeq & e^{i (q \phi - \mu t)} \left[ \tilde \psi_0 +
\varepsilon_n \tilde u_n e^{i (\delta m \phi - E_n t)} + \varepsilon_n^*
\tilde v_n^* e^{-i (\delta m \phi - E_n^* t)} \right] \nonumber \\
& = & e^{i (q \phi - \mu t)} \Bigl\{ \tilde \psi_0 + e^{{\rm Im}E_n t}
|\varepsilon_n| \Bigl[ (\tilde u_n + \tilde v_n) \cos(\delta m \phi - {\rm
Re}E_n t + \theta_n) + (\tilde u_n- \tilde v_n) \sin(\delta m \phi - {\rm
Re}E_n t + \theta_n) \Bigr] \Bigr\},
\end{eqnarray}
\end{widetext}
where we leave only the Bogoliubov mode with the largest imaginary part
${\rm Im}E_n > 0$, and define $\psi_0 \equiv \tilde \psi_0 e^{i q \phi} $,
$u_n \equiv \tilde u_n e^{i (q + \delta m) \phi}$, $v_n \equiv \tilde v_n
e^{-i (q - \delta m) \phi}$, and $\varepsilon_n \equiv |\varepsilon_n|
e^{i \theta_n}$.
Since both $(\tilde u, \tilde v, E)$ and $(\tilde u^*, \tilde v^*, E^*)$
are solutions of Eq.~(\ref{Bogo}), one of them satisfies ${\rm Im}E_n >
0$.
Equation~(\ref{modegrow}) indicates that the $\delta m$-fold density
modulation grows exponentially in time while rotating at frequency ${\rm
Re}E_n / \delta m$.
Note, however, that the above discussion, which is based on Bogoliubov
analysis, can be applied only to an initial stage of the growth of the
unstable mode, and we must integrate GP equation (\ref{GPN}) to follow the
time evolution over a longer time.

We briefly describe the numerical procedure which will be used to solve
Eqs.~(\ref{GPN}) and (\ref{Bogo}).
The stationary state $\psi_0$ of Eq.~(\ref{GPN}) is obtained by the
Newton-Raphson method~\cite{Edwards} or the imaginary time
method~\cite{Dalfovo} in the restricted functional subspace of $\psi_0
\propto e^{i q \phi}$.
The Bogoliubov spectra are then obtained by diagonalizing
Eq.~(\ref{Bogo}) in each subspace of the angular momentum $\delta m$.
Time evolution of the wave function is obtained by integrating
Eq.~(\ref{GPN}) using the Crank-Nicholson scheme~\cite{Ruprecht} with a
spatial discretization taken typically to be $256 \times 256 \times 256$.
In studying dynamical instabilities, we add a symmetry-breaking
perturbation to the initial state such as
\begin{equation} \label{perturb}
\psi(t = 0) \propto (1 + 0.01 \cos n \phi) \psi_0,
\end{equation}
where $n$ is an integer.
While dynamically unstable modes can grow from infinitesimal numerical
noise over a sufficiently long time even when we start from an
axisymmetric state, we add the perturbation (\ref{perturb}) in order to
simulate experimental noise in a controllable manner so that the results
are reproducible.
This type of perturbation may be realized experimentally, e.g., using
off-resonant laser beams propagating along the trap axis.
The system is allowed to evolve in time until just before the collapse
sets in, since numerical simulations of violent collapses and
explosions~\cite{Donley,SaitoLA} in 3D space present a formidable
computational challenge.

\section{Dynamical instabilities in a singly-quantized vortex}
\label{s:single}

\begin{figure}[tb]
\includegraphics[width=8.4cm]{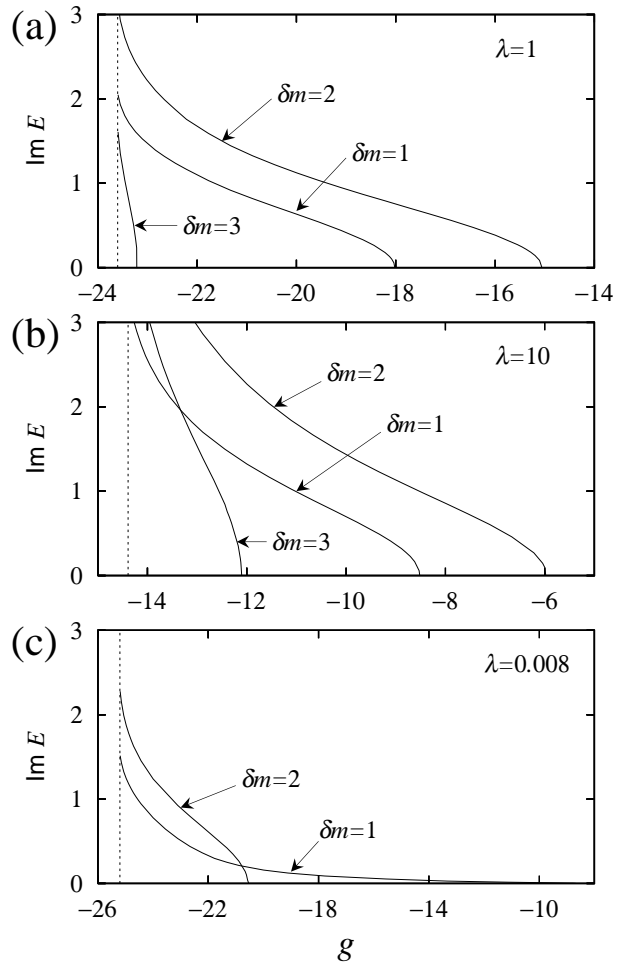}
\caption{
Imaginary parts of the Bogoliubov spectra of the singly-quantized
vortex state as functions of $g$ for (a) $\lambda = 1$ (spherical trap),
(b) $\lambda = 10$ (pancake-shaped), and (c) $\lambda = 0.008$
(cigar-shaped).
The dotted lines show critical values $g_M^{\rm cr}$ below which the
collapse of the vortex state is caused by the monopole instability.
}
\label{f:L1}
\end{figure}

In this section, we study the dynamical instabilities in a
singly-quantized vortex state $\psi_0 \propto e^{i\phi}$.
Figure~\ref{f:L1} shows imaginary parts of the Bogoliubov spectra as
functions of the interaction parameter $g$ for spherical ($\lambda = 1$),
pancake-shaped ($\lambda = 10$), and cigar-shaped ($\lambda =
0.008$~\cite{Chin}) traps.
As seen from Figs.~\ref{f:L1} (a) and (b), for the spherical and
pancake-shaped traps, the quadrupole mode ($\delta m = 2$) first becomes
unstable in the sense that the imaginary part ${\rm Im}E$ appears for
values of $|g|$ smaller than those of other modes.
A similar result is obtained in Ref.~\cite{Pu} for a 2D system, which
corresponds to the pancake-shaped trap with a large asymmetry parameter
$\lambda$~\cite{Castin}.
For the cigar-shaped trap [Fig.~\ref{f:L1} (c)], on the other hand, the
dipole mode ($\delta m = 1$) first becomes unstable, which occurs for
$\lambda \lesssim 0.34$~\cite{SaitoL}.
This suggests that when the strength of interaction $|g|$ is adiabatically
increased, the dipole instability plays a dominant role of decay
mechanisms for the cigar-shaped trap with $\lambda \lesssim 0.34$;
otherwise the quadrupole mode is dominant.

\begin{figure}[tb]
\includegraphics[width=8.4cm]{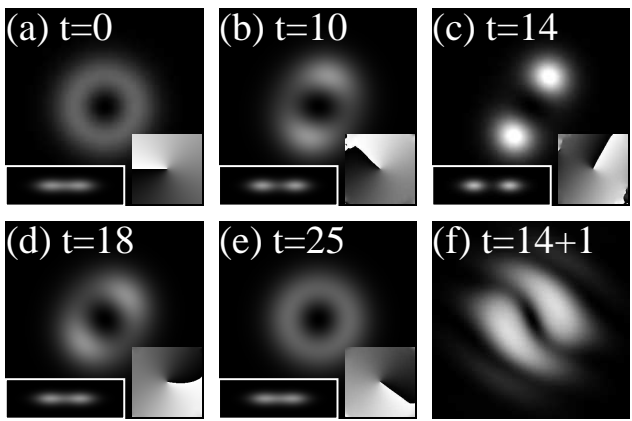}
\caption{
(a)-(e) Snapshots of integrated densities $\rho_z(x, y) = \int dz
|\psi|^2$ (main panels) and $\rho_x(y, z) = \int dx |\psi|^2$ (lower-left
insets) for $g = -6.5$ and $\lambda = 10$.
The initial state is the stationary singly-quantized vortex state for $g =
-6.5$ disturbed by perturbation (\protect\ref{perturb}) with $n = 2$.
The lower-right insets show gray-scale images of the phase of the
integrated wave function $\Psi_z = \int dz \psi$,
where the phase from $(2 n - 1) \pi$ to $(2 n + 1) \pi$ is represented in
gray scale from black to white.
(f) The integrated density $\rho_z$ at $t = 15$, where the interaction is
switched to $g = 50$ and the trap is turned off at $t = 14$.
The sensitivity of the imaging in (f) is 12 times higher than that in
(a)-(e).
The sizes of the images are $6 \times 6$ in $\rho_z$ and the phase
profiles, and $6 \times 2$ in $\rho_x$ in (a)-(e), and $14 \times 14$ in
(f) in units of $(\hbar / m \omega_\perp)^{1/2}$.
}
\label{f:panL1}
\end{figure}

Figure~\ref{f:panL1} shows the time evolution of integrated column
densities and phase profiles for $\lambda = 10$.
The integrated column density is defined by
\begin{equation}
\rho_\xi \equiv \int d\xi |\psi|^2 \;\;\; (\xi = x, y, z),
\end{equation}
and the phase of the integrated wave function $\int dz \psi$ is used to
draw the phase profiles.
The main panels in Fig.~\ref{f:panL1} show top views $\rho_z$ and the
lower-left insets show side views $\rho_x$.
The interaction parameter is taken to be $g = -6.5$, which exceeds the
critical value for the quadrupole instability $g_Q^{\rm cr} = -6.0$, but
is below the critical value for the dipole instability $g_D^{\rm cr} =
-8.5$.
The Bogoliubov eigenvalue of the quadrupole mode at $g = -6.5$ is $E_Q
\simeq 1.5 + 0.40 i$.
We add the perturbation~(\ref{perturb}) with $n = 2$ to the initial state
to trigger the quadrupolar dynamical instability.
In Fig.~\ref{f:panL1} (b), we see that a density fluctuation in the
quadrupole mode starts to grow with the Lyapunov exponent ${\rm Im} E_Q
\simeq 0.4$ and rotates at frequency ${\rm Re} E_Q / 2 \simeq 0.75$ in
accordance with Eq.~(\ref{modegrow}).
The vortex then splits into two clusters which revolve around the center
of the trap [Fig.~\ref{f:panL1} (c)], and subsequently these clusters
unite to recover the original vortex [Fig.~\ref{f:panL1} (e)].
After this, the split-merge cycle repeats.
Figure~\ref{f:panL1} (f) shows an expanded image of $\rho_z$ at $t = 15$,
in which the interaction was switched from $g = -6.5$ to $g = 50$ and the
trap was turned off at $t = 14$.
The fringe pattern is due to interference between two matter waves
emanating from the two clusters shown in Fig.~\ref{f:panL1} (c).
The above results confirm that our findings in Ref.~\cite{SaitoL}, which
were obtained in a 2D system, hold true in a pancake-shaped trap and are
therefore realizable experimentally.

When the vortex splits into $n$ clusters, each of these subsequently
collapses if the effective strength of interaction per each cluster, i.e.,
$|g| / n$, exceeds the critical value $|g_{\rm nonvortex}^{\rm cr}|$ for
the collapse of a condensate without vortices, while the split-merge cycle
occurs if $|g| / n$ is sufficiently smaller than $|g_{\rm nonvortex}^{\rm
cr}|$.
In fact, in the case of $\lambda = 10$ in Fig.~\ref{f:panL1}, $|g| / 2 =
3.25$ is smaller than $|g_{\rm nonvortex}^{\rm cr}| = 3.73$, and therefore
the split clusters do not collapse, allowing the system to undergo
split-merge cycles.
However, for the spherical trap, $g_{\rm nonvortex}^{\rm cr} = -7.22$ and
$g_Q^{\rm cr} = -15.1$, and hence the value $|g_Q^{\rm cr}| / 2 = 7.55$ at
which the split occurs is always larger than $|g_{\rm nonvortex}^{\rm
cr}|$, indicating that split clusters always collapse.
We performed numerical calculations for the spherical trap with $|g|$
slightly larger than $|g_Q^{\rm cr}|$, and confirmed that two clusters
immediately collapse without undergoing a split-merge cycle (as shown in
Figs.~4 (d)-(f) of Ref.~\cite{SaitoL}).
The phenomenon, in which a vortex first splits into fragments and each
fragment subsequently collapses, may be called ``fragmented collapse''.

\begin{figure}[tb]
\includegraphics[width=8.4cm]{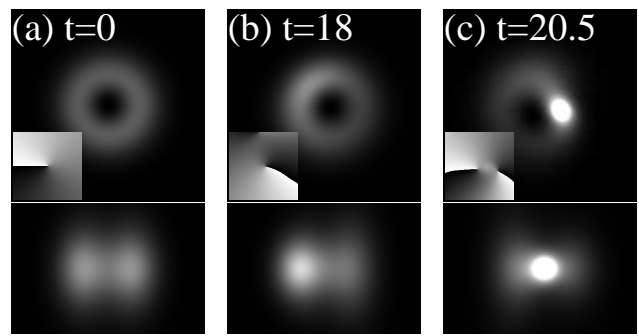}
\caption{
Snapshots of integrated densities $\rho_z(x, y)$ (upper panels),
$\rho_x(y, z)$ (lower panels), and phase profiles (insets) for $g = -20$
and $\lambda = 0.008$.
The initial state is the stationary singly-quantized vortex state for $g =
-20$ disturbed by perturbation (\protect\ref{perturb}) with $n = 1$.
The sizes of the images are $6 \times 6$ for $\rho_z$ and $6 \times 4$ for
$\rho_x$ in units of $(\hbar / m \omega_\perp)^{1/2}$.
}
\label{f:mitL1}
\end{figure}

Figure~\ref{f:mitL1} shows time evolution for the cigar-shaped trap with
$\lambda = 0.008$.
The interaction parameter is taken to be $g = -20$, at which only the
dipole mode is unstable [see Fig.~\ref{f:L1} (c)].
The initial state is the stationary state for $g = -20$ plus the small
perturbation (\ref{perturb}) with $n = 1$.
We see that the atoms tend towards one side of the ring due to the dipole
instability [Fig.~\ref{f:mitL1} (b)], after which collapse occurs
[Fig.~\ref{f:mitL1} (c)].
Since the dipole mode has angular momenta $m = 1 \pm 1 = 2$ and $0$, the
phase defect that exists from the outset slightly deviates from the center
associated with the growth of the $m = 0$ component, and another phase
defect enters due to the $m = 2$ component as seen in the inset of
Fig.~\ref{f:mitL1} (c).
For $\lambda = 0.008$, the critical value at which the dipole mode becomes
unstable is $g_D^{\rm cr} = -8.66$, which exceeds $g_{\rm nonvortex}^{\rm
cr} = -8.49$, and therefore the dipole instability is always followed by
collapse.
We note that the dipole instability in Fig.~\ref{f:mitL1} is distinct from
the dissipative vortex spiral-out phenomenon that arises from the dipole
mode with negative eigenvalue~\cite{Rokhsar,Isoshima,Fetter98}.
In the former, the quasiparticles that change angular momentum by $\pm 1$
are excited in pairs, and hence the angular momentum is conserved.
The energy is also conserved, since the real parts of the excitation
energies have the same magnitude with opposite signs.
In the latter, the atoms are transferred only into the non-vortex ground
state decreasing the angular momentum and dissipating the energy.

\section{Dynamical instabilities in a doubly-quantized vortex}
\label{s:double}

A doubly-quantized vortex state $\propto e^{2 i \phi}$ of ${}^{23}{\rm
Na}$ BEC has recently been realized by the MIT group~\cite{Leanhardt}
using a topological phase-imprinting method~\cite{Nakahara}, where the
interaction is repulsive.
The doubly-quantized vortex with repulsive interactions is predicted to be
dynamically unstable against disintegration into two singly-quantized
vortices~\cite{Nozieres,Garcia99,Mottonen}.
In the case of attractive interactions, we will show that not only the
vortex disintegration but also various new phenomena, such as a three-fold
split-merge process, occur due to dynamical instabilities.

\begin{figure}[tb]
\includegraphics[width=8.4cm]{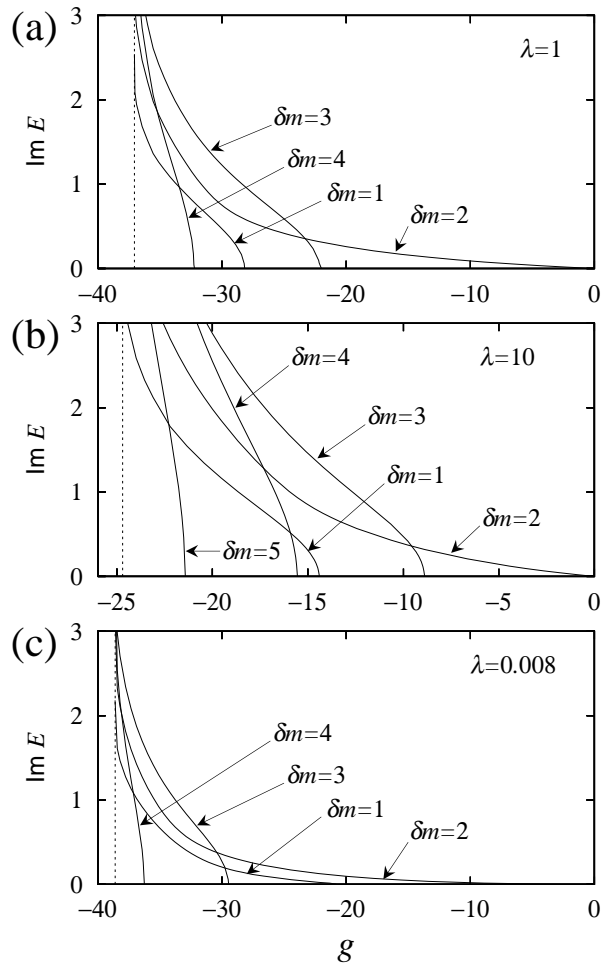}
\caption{
Imaginary parts of the Bogoliubov spectra of the doubly-quantized
vortex state as functions of $g$ for (a) $\lambda = 1$, (b) $\lambda =
10$, and (c) $\lambda = 0.008$.
The dotted lines show the critical value $g_M^{\rm cr}$ below which the
collapse of the vortex state is caused by the monopole instability.
}
\label{f:L2}
\end{figure}

The imaginary parts of the Bogoliubov eigenvalues for the doubly-quantized
vortex states are shown in Fig.~\ref{f:L2}.
We note that unlike the case of the singly-quantized vortex state, the
quadrupole mode ($\delta m = 2$) is always dynamically unstable for $g <
0$, indicating that no dynamically stable state exists for the case of
attractive interactions.
Therefore, an adiabatic decrease in $g$ from positive to negative always
gives rise to the quadrupole instability.

\begin{figure}[tb]
\includegraphics[width=8.4cm]{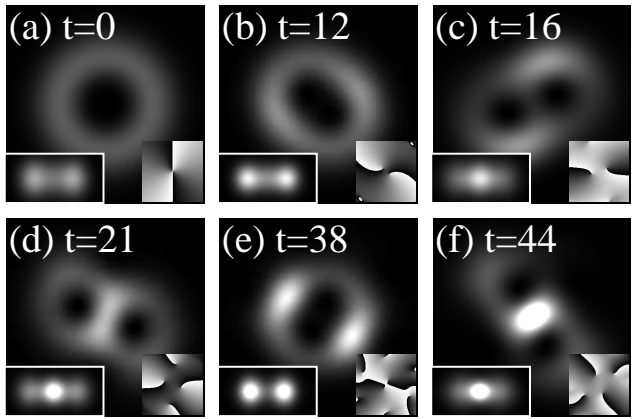}
\caption{
Snapshots of integrated densities $\rho_z(x, y)$ (main panels), $\rho_y(x,
z)$ (lower-left insets), and phase profiles (lower-right insets) for
$\lambda = 1$.
The initial state is the stationary doubly-quantized vortex state for $g =
0$ disturbed by perturbation (\protect\ref{perturb}) with $n = 2$, and the
interaction is switched to $g = -20$ at $t = 0$.
The sizes of the images are $6 \times 6$ for $\rho_z$ and the phase
profiles, and $6 \times 3$ for $\rho_y$ in units of $(\hbar / m
\omega_\perp)^{1/2}$.
}
\label{f:sphL2}
\end{figure}

Figure~\ref{f:sphL2} shows time evolutions of the integrated densities
$\rho_z$ (main panels) and $\rho_y$ (lower-left insets) and the phase
profiles (lower-right insets) for the doubly-quantized vortex state in a
spherical trap, where the initial state is a stationary state of the GP
equation with $g = -20$ plus the perturbation given in Eq.~(\ref{perturb})
with $n = 2$.
For this strength of interaction, only the quadrupole mode is dynamically
unstable [cf. Fig.~\ref{f:L2} (a)].
The vortex core disintegrates, as can be seen from the density and phase
profiles.
The angular momenta of the quadrupole excitations are $m = 2 \pm 2 = 4$
and $0$.
Vortex disintegration is attributed to the growth of the $m = 0$
component, and the four phase defects in Fig.~\ref{f:sphL2} (d) are due to
the $m = 4$ component.
The density distribution exhibits various changes such as high-density
regions occurring in the opposite edge of the vortices [Fig.~\ref{f:sphL2}
(c)], then between the two vortices [Fig.~\ref{f:sphL2} (d)], and then the
density distribution becomes similar to the split vortex as in the case of
a singly-quantized vortex [Fig.~\ref{f:sphL2} (e)].
Finally, in Fig.~\ref{f:sphL2} (f) the central density exceeds critical
density and collapses.

\begin{figure}[tb]
\includegraphics[width=8.4cm]{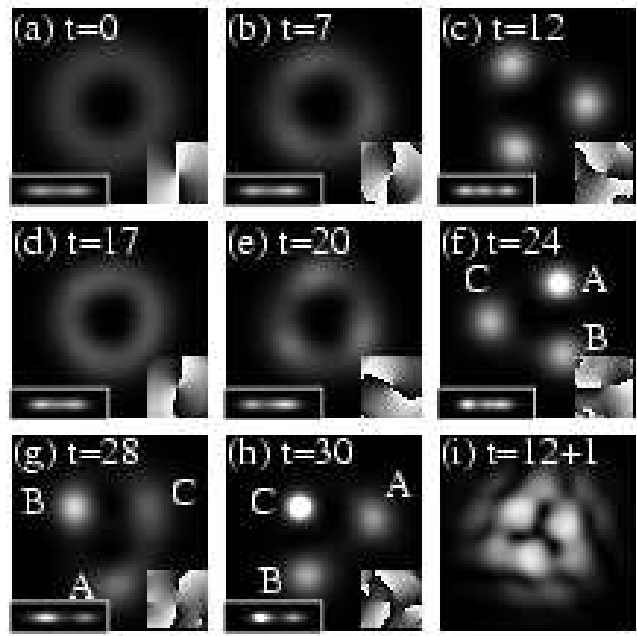}
\caption{
(a)-(h) Snapshots of integrated densities $\rho_z(x, y)$ (main panels),
$\rho_x(y, z)$ (lower-left insets), and phase profiles (lower-right
insets) for the pancake-shaped trap with $\lambda = 10$.
The initial state is the stationary doubly-quantized vortex state for $g =
0$ disturbed by perturbation (\protect\ref{perturb}) with $n = 3$, and the
interaction is switched to $g = -10$ at $t = 0$.
(i) The integrated density $\rho_z$ at $t = 13$, where the interaction is
switched to $g = 50$ and the trap is turned off at $t = 12$.
The sensitivity of the imaging is 12 times higher than that in (a)-(h).
The sizes of the images are $6 \times 6$ in $\rho_z$ and the phase
profiles, and $6 \times 2$ in $\rho_x$ in (a)-(h), and $14 \times 14$ in
(i) in units of $(\hbar / m \omega_\perp)^{1/2}$.
}
\label{f:panL2}
\end{figure}

Another important distinction of Fig.~\ref{f:L2} from Fig.~\ref{f:L1} is
that there is a range of $g$ over which the imaginary part of the
eigenvalue of the octupole mode ($\delta m = 3$) becomes largest.
This implies that if $g$ is suddenly changed into that regime, the
octupolar instability dominates until the quadrupole mode grows.
Figure~\ref{f:panL2} shows a typical example of such situations with
$\lambda = 10$, where the initial state is the noninteracting stationary
state with vorticity of two subject to perturbation~(\ref{perturb}) with
$n = 3$, and the interaction parameter is changed from $g = 0$ to $g =
-10$ at $t = 0$.
As the octupole mode grows, the vortex splits into three clusters which
revolve around the center of the trap [Fig.~\ref{f:panL2} (c)].
Interestingly, the three clusters merge to recover the original vortex
[Fig.~\ref{f:panL2} (d)] as in the two-cluster case shown in
Fig.~\ref{f:panL1}.
This is because the effective strength of interaction per each cluster
$|g| / 3 = 3.33$ is smaller than the critical value for collapse of the
non-vortex state $|g^{\rm cr}_{\rm nonvortex}| = 3.73$.
Subsequently, the vortex once again splits into three clusters, and the
balance in atomic numbers between these is then broken by the quadrupole
instability.
It follows that one cluster [denoted by A in Fig.~\ref{f:panL2} (f)]
grows, then another cluster [B in Fig.~\ref{f:panL2} (g)] grows, and
finally the remaining cluster [C in Fig.~\ref{f:panL2} (h)] grows and
eventually collapses.
This phenomenon is similar to the seesaw-like oscillation between two
clusters due to the dipole instability~\cite{SaitoL}.
At the initial stage of disintegration in which the octupole mode plays a
dominant role [Figs.~\ref{f:panL2} (a)-(e)], the split in the phase
defects is not very prominent because the angular momenta of the $\delta
m = \pm 3$ modes are $m = 2 \pm 3 = 5$ and $-1$.
When the quadrupole mode becomes significant, the outward shifts of
the phase defects manifest themselves [Figs.~\ref{f:panL2} (g)-(h)] due to
the $m = 0$ component of the quadrupole mode.
Figure~\ref{f:panL2} (i) shows the integrated density at $t = 13$,
following the interaction being switched from $g = -10$ to $g = 50$ and
the trap being turned off at $t = 12$.
We see the three-fold fringe pattern resulting from interference between
the matter waves that emanate from the three clusters shown in
Fig.~\ref{f:panL2} (c).

\begin{figure}[tb]
\includegraphics[width=8.4cm]{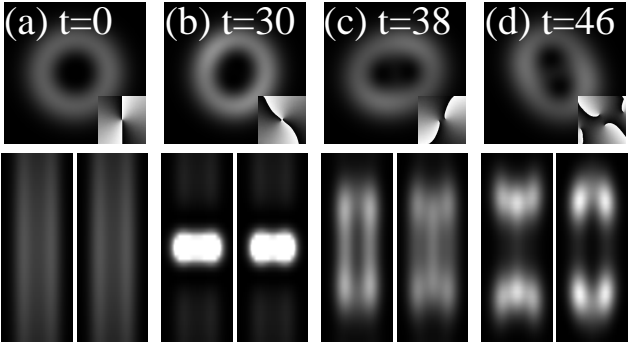}
\caption{
Snapshots of $\rho_z(x, y)$ (upper panels), phase profiles (insets),
$\rho_x(y, z)$ (lower-left panels), and  $\rho_y(x, z)$ (lower-right
panels) for the cigar-shaped trap with $\lambda = 0.008$.
The initial state is the stationary doubly-quantized vortex state for $g =
0$ disturbed by perturbation (\protect\ref{perturb}) with $n = 2$, and the
interaction is switched to $g = -38$ at $t = 0$.
The sizes of the images are $6 \times 6$ for $\rho_z$ and the phase
profiles, and $6 \times 16$ for $\rho_x$ and $\rho_y$ in units of $(\hbar
/ m \omega_\perp)^{1/2}$.
}
\label{f:mitL2}
\end{figure}

In an elongated cigar-shaped trap, dynamics along the trap axis also plays
an important role.
Figure~\ref{f:mitL2} shows the integrated density and phase profiles for
$\lambda = 0.008$.
The initial state is the noninteracting stationary state with vorticity of
two plus perturbation (\ref{perturb}) with $n = 2$, and the interaction is
switched from $g = 0$ to $g = -38$ at $t = 0$.
The atoms first gather along the symmetry axis (i.e., the $z$ axis)
towards $z = 0$ [Fig.~\ref{f:mitL2}(b)], then the vortex begins to
disintegrate since the density is high enough.
Subsequently the atoms rebound along the $z$ direction
[Fig.~\ref{f:mitL2}(c)].
We see that the disintegrated vortex lines in Fig.~\ref{f:mitL2} (c) are
straight, unlike the repulsive case where twisting of vortex lines
occurs~\cite{Mottonen}.
In Fig.~\ref{f:mitL2} (d), the condensate splits into two high-density
regions containing the disintegrated vortices as a consequence of the
rebound.

\begin{figure}[tb]
\includegraphics[width=8.4cm]{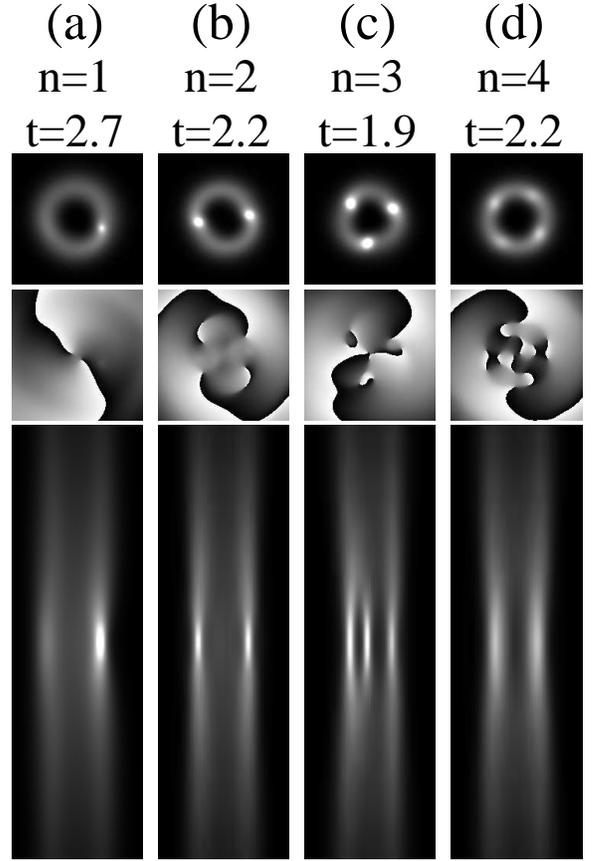}
\caption{
Snapshots of $\rho_z(x, y)$, phase profiles, and $\rho_y(x, z)$ (from top
to bottom) for the cigar-shaped trap with $\lambda = 0.008$.
The initial state is the stationary doubly-quantized vortex state for $g =
0$ disturbed by perturbation (\protect\ref{perturb}) with (a) $n = 1$, (b)
$n = 2$, (c) $n = 3$, and (d) $n = 4$.
The interaction is switched from $g = 0$ to $g = -500$ at $t = 0$.
The sizes of the images are $6 \times 6$ for $\rho_z$ and the phase
profiles, and $6 \times 20$ for $\rho_y$ in units of $(\hbar / m
\omega_\perp)^{1/2}$.
}
\label{f:mitL2exp}
\end{figure}

In an experiment using ${}^{23}{\rm Na}$, the lifetime of the condensate
is as short as 1 or 2 ms near the Feshbach resonance~\cite{Jitkee}, which
indicates that the lifetime is $t \lesssim 3$ (in our dimensionless units
of time) for the optical dipole trap used in the MIT group ($\omega_\perp
= 2\pi \times 250$ Hz, $\omega_z = 2\pi \times 2$ Hz, and $\lambda =
0.008$)~\cite{Chin}.
For dynamical instabilities to occur within this short lifetime, the
strength of attractive interaction $|g|$ must be much larger than those in
the above cases.
Figure~\ref{f:mitL2exp} shows the density and phase profiles just before
the collapse, where the initial state is the noninteracting stationary
state with vorticity of two plus perturbation (\ref{perturb}) with $n =
1$, $2$, $3$, and $4$, and the interaction is switched from $g = 0$ to $g
= -500$ at $t = 0$.
We find that the collapsing dynamics depend on the initial perturbation
$n$, since the system has several multipole instabilities for $g = -500$.
We note that the vortex split occurs only around $z = 0$, where the
density is high enough.
Thus, the vortex-split phenomena can be observed for large values of
$|g|$, even when the lifetime of the condensate is very short.

\section{Variational analysis}
\label{s:variational}

To understand the split-merge phenomena shown in the preceding sections,
we restrict ourselves to 2D space in this and next sections.

We employ a Gaussian variational wave function for the condensate as
\begin{equation} \label{Gaussian0}
\psi_q = \frac{r^{|q|}}{\sqrt{\pi |q|!} d^{|q| + 1}} e^{-\frac{r^2}{2d^2} 
+ i q \phi},
\end{equation}
where $q$ is the vorticity and $d$ is a variational parameter
characterizing the size of the condensate.
Substituting Eq.~(\ref{Gaussian0}) into the GP energy functional
\begin{equation} \label{E}
E[\psi] = \int d{\bf r} \left[ \frac{1}{2} \psi^* \left( -\nabla^2 + 
\omega^2 r^2 \right) \psi + \frac{g}{2} |\psi|^4 \right]
\end{equation}
and minimizing it with respect to $d$ gives
\begin{equation} \label{d}
d = \left[ 1 + \frac{(2 |q|)! g}{2^{2 |q| + 1} |q|!^2 (1 + |q|) \pi}
\right]^{\frac{1}{4}}.
\end{equation}
For $g < g_M^{\rm cr} \equiv -2^{2 |q| + 1} |q|!^2 (1 + |q|) \pi / (2
|q|)!$, the energy functional (\ref{E}) has no extremum, and the system
collapses due to the monopole instability.
The critical values $g_M^{\rm cr} = -8\pi$ for $q = 1$ and $g_M^{\rm cr} =
-16\pi$ for $q = 2$ agree reasonably well with the numerically obtained
critical values $-24$ and $-45$, respectively.

To study the time evolution from Eq.~(\ref{Gaussian0}), we use a
variational wave function given by
\begin{equation} \label{varpsi}
\psi = \sum_m c_m \psi_m,
\end{equation}
where $\sum_m |c_m|^2 = 1$ and $\psi_m$ is given by Eq.~(\ref{Gaussian0})
with $q$ replaced by $m$.
In Eq.~(\ref{varpsi}) we use the same width $d$ as Eq.~(\ref{d}) for all
$\psi_m$ and neglect the radial degrees of freedom to simplify the
following calculation and to obtain the results analytically.

Substituting Eq.~(\ref{varpsi}) into the action
\begin{equation}
S[\psi] = \int dt E[\psi] - i \int d{\bf r} dt \psi^* 
\frac{\partial}{\partial t} \psi
\end{equation}
yields
\begin{eqnarray} \label{S}
S & = & \int dt \Biggl( -i \sum_m c_m^* \dot c_m + \sum_m \varepsilon_m 
|c_m|^2 \nonumber \\
& & + \frac{g}{2} \sum_{\stackrel{\scriptstyle m_1 m_2}{m_3 m_4}}
G^{m_1, m_2}_{m_3, m_4} c_{m_1}^* c_{m_2}^* c_{m_3} c_{m_4} \Biggr),
\end{eqnarray}
where $\varepsilon_m \equiv (1 + |m|) (d^2 + 1 / d^2) / 2$, and $G^{m_1,
m_2}_{m_3, m_4} \equiv \int d{\bf r} \psi_{m_1}^* \psi_{m_2}^* \psi_{m_3} 
\psi_{m_4}$, which vanishes for $m_1 + m_2 - m_3 - m_4 \neq 0$.
From $\partial S / \partial c_m^* = 0$, the variational equation of motion
for $c_m$ becomes
\begin{equation} \label{eomc}
i \dot c_m = \varepsilon_m c_m + g \sum_{m_1 m_2 m_3} G^{m, m_1}_{m_2,
m_3} c_{m_1}^* c_{m_2} c_{m_3}.
\end{equation}
We assume $|c_q| \simeq 1$ and $|c_{m \neq q}| \ll 1$, and neglect terms
of the order of $O(|c_{m \neq q}|^2)$.
Equation~(\ref{eomc}) for $m = q$ reduces to
\begin{equation} \label{c1}
i \dot c_q \simeq (\varepsilon_1 + \frac{(2 |q|)! g}{2^{2 |q| + 1} |q|!^2
\pi d^2}) c_q \equiv \mu c_q,
\end{equation}
and hence we obtain $c_q \simeq e^{-i \mu t}$.
Using Eq.~(\ref{c1}) and defining $c_m \equiv e^{-i\mu t} \tilde c_m$, we
obtain a closed set of equations
\begin{subequations}
\begin{eqnarray}
i \dot{\tilde c}_m & \simeq & \left( \varepsilon_m - \mu + 2g G^{m, q}_{m,
q} \right) \tilde c_m \nonumber \\
& & + g G^{m, 2q - m}_{q, q} \tilde c_{2q - m}^*, \\
-i \dot{\tilde c}_{2q - m}^* & \simeq & \left( \varepsilon_{2q - m} - \mu
+ 2g G^{2q - m, q}_{2q - m, q} \right) \tilde c_{2q - m}^* \nonumber \\
& & + g G^{2q - m, m}_{q, q} \tilde c_m,
\end{eqnarray}
\end{subequations}
from which the eigenfrequencies can be found.
The condition under which the eigenfrequencies are real is given by
\begin{eqnarray}
& & D_{q, m} = D_{q, 2q - m} \nonumber \\
& & \equiv \left[ \varepsilon_m + \varepsilon_{2q
- m} - 2\mu + 2g \left( G^{m, q}_{m, q} + G^{2q - m, q}_{2q - m, q} \right)
\right]^2 \nonumber \\
& & - 4g^2 \left( G^{m, 2q - m}_{q, q} \right)^2 \geq 0.
\end{eqnarray}

First we consider the singly-quantized vortex ($q = 1$).
Stability of the quadrupole mode is determined by
\begin{equation} \label{D13}
D_{1, 3} = 3 + \frac{5g}{8\pi} + \frac{1}{1 + \frac{g}{8\pi}} \left( 1 +
\frac{g}{2\pi} - \frac{g^2}{32\pi^2} \right),
\end{equation}
which becomes negative for $g < -9.2$.
This value is in reasonable agreement with the numerically obtained
critical value for the quadrupole mode in 2D, $g_Q^{\rm cr} = -7.8$.
However, the discriminant $D_{1, 2} = g^2 / (64 \pi^2 d^4)$ is always
positive, which contradicts the numerical result that dipole instability
does occur for $g < g_D^{\rm cr} = -11.5$.
Also, for $m = -2$ or $m = 4$, the numerical calculation shows that
dynamical instability occurs for $g < -16.9$, while $D_{1, 4}$ is always
positive for $g > -8\pi$ [$g$ must be larger than $-8\pi$ as seen from
Eq.~(\ref{d})].
In the case of the doubly-quantized vortex ($q = 2$), $D_{2, 4} = -47 g^2
/ (4096 \pi^2 d^4)$ is always negative, which is consistent with the fact
that the quadrupole mode is always dynamically unstable for $g < 0$.
However, $D_{2, 4}$ fails to reproduce the behavior for $g > 0$, where
stable and unstable regions appear alternatively~\cite{Pu,Mottonen}.
The discriminant $D_{2, 5}$ becomes negative for $g < -15.8$, which is in
reasonable agreement with the numerical result that the octupole
instability sets in for $g < -11.7$.
However, $D_{2, 3} = g^2 / (256 \pi^2 d^2) \geq 0$ contradicts the
numerical result that dipole instability does occur for $g < g_D^{\rm cr}
= -19.8$.

Thus, the Gaussian ansatz (\ref{varpsi}) is viable for some cases, whereas
more sophisticated trial functions are needed to completely describe the
dynamical instabilities.
The variational results can be improved by using the different variational
width $d$ for each $\psi_m$ in Eq.~(\ref{varpsi}), which may be determined
by, e.g., the method used in Ref.~\cite{Fetter}.
However, near and above the critical point, this variational method fails,
since the norm of the Bogoliubov function vanishes when a complex
eigenvalue emerges.

Equation~(\ref{D13}) is expanded near the critical point as
\begin{equation}
D_{1, 3} = \frac{3 \sqrt{3}}{(4 \sqrt{3} - 3 \sqrt{2})\pi} (g - g^{\rm
cr}) + O\left( (g - g^{\rm cr})^2 \right),
\end{equation}
where $g^{\rm cr} = 16 (\sqrt{6} - 3) \pi / 3$.
Just above the critical point, therefore, the imaginary part of the
complex eigenvalue is proportional to $(-D_{1, 3})^{1/2} \propto (g^{\rm
cr} - g)^{1/2}$~\cite{SaitoL}.
All the imaginary parts shown in Figs.~\ref{f:L1} and \ref{f:L2} (except
$\delta m = 2$ in Fig.~\ref{f:L2}) have this $g$ dependence near the
critical points [so do the $\delta m = 1$ curves in Figs.~\ref{f:L1} (c)
and \ref{f:L2} (c) in the immediate vicinity of the critical points].
The $\delta m = 2$ lines for $q = 2$ in Fig.~\ref{f:L2} are proportional
to $|g|$ near $g = 0$, which agrees with $(-D_{2, 4})^{1/2} \propto |g|$.

\begin{figure}[tb]
\includegraphics[width=8.4cm]{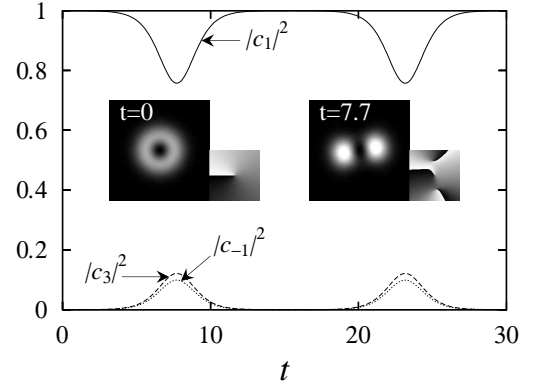}
\caption{
Time evolutions of $c_1$ (solid curve), $c_3$ (dashed curve), and $c_{-1}$
(dotted curve) in the three-mode approximation of Eq.~(\protect\ref{eomc})
for $g = -12$.
The initial condition is $c_3 = c_{-1} = 0.01$ and $c_1 = (1 - 2 \cdot
0.01^2)^{1/2}$.
The insets show the density and phase profiles at $t = 0$ and $t = 7.7$,
where the sizes of the images are $6 \times 6$ in units of $(\hbar / m
\omega)^{1/2}$.
}
\label{f:3mode}
\end{figure}

The split-merge cycles shown in Figs.~\ref{f:panL1} and \ref{f:panL2} are
recurrence phenomena arising from nonlinearity.
To understand the behavior in Fig.~\ref{f:panL1} qualitatively, we reduce
the number of modes in Eq.~(\ref{eomc}) to three, i.e., $m = 1$, $3$, and
$-1$.
Equation~(\ref{eomc}) then reduces to simultaneous nonlinear
differential equations with three variables, which can be solved
numerically.
Figure~\ref{f:3mode} shows the time evolution of $|c_1|^2$, $|c_3|^2$, and
$|c_{-1}|^2$ for $g = -12$, where the initial condition is $c_3 = c_{-1} =
0.01$ and $c_1 = (1 - 2 \cdot 0.01^2)^{1/2}$.
The vortex can be seen to split at $t \simeq 7.7$, but subsequently
recovers the original vortex during $12 \lesssim t \lesssim 18$, and then
splits again at $t \simeq 23.2$. 
Thus, the split-merge phenomenon can be qualitatively reproduced by a
simple three-mode approximation.

\section{Low-lying soliton state and Goldstone mode}
\label{s:lowly}

The split-merge dynamics shown in Fig.~\ref{f:panL1} implies the existence
of a low-lying state in which two clusters revolve around each other
without changing their shapes.
To find such a low-lying stationary state, we numerically seek a local 
minimum of the GP energy functional with the angular momentum held fixed
at that of the initial vortex state $\langle \hat L_z \rangle = 1$,
where $\hat L_z = -i \partial / \partial \phi$.
The GP energy functional with Lagrange multipliers is given by
\begin{equation}
F[\psi] = E[\psi] + \int d{\bf r} \left( -\mu |\psi|^2 + \Omega \psi^* 
\hat L_z \psi \right),
\end{equation}
where $E[\psi]$ is given in Eq.~(\ref{E}), and the chemical potential
$\mu$ and the angular frequency $\Omega$ of rotation of the system are
introduced as the Lagrange multipliers in order to fix the normalization
and the angular momentum.

\begin{figure}[tb]
\includegraphics[width=7cm]{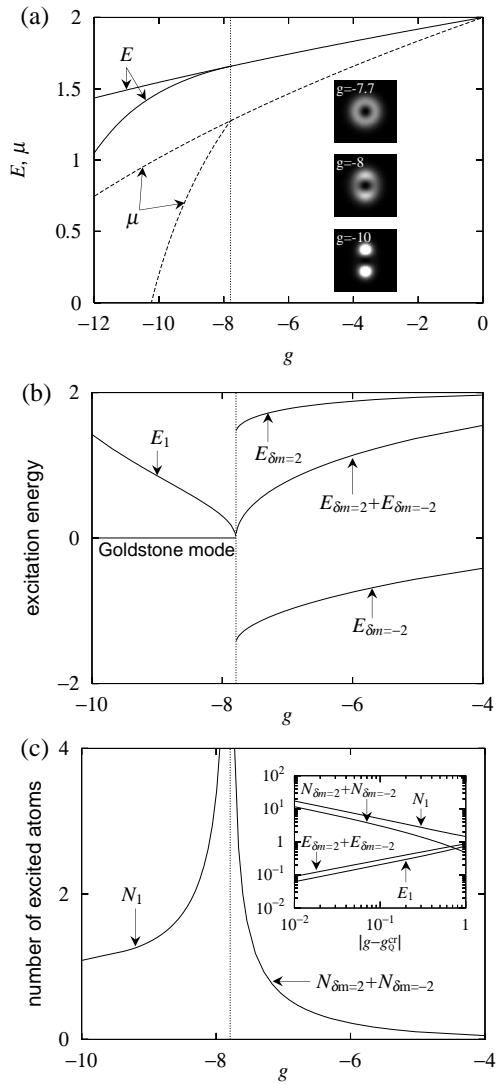}
\caption{
(a) The energy $E$ (solid curves) and chemical potential $\mu$ (dashed
curves) of the axisymmetric singly-quantized vortex state (upper branch)
and low-lying soliton state (lower branch for $g < -7.79$).
Gray-scale images show the density profile of the axisymmetric vortex
state for $g = -7.7$ and those of the low-lying states for $g = -8$ and $g
= -10$.
(b) The Bogoliubov excitation energies of the quadrupole modes above the
axisymmetric vortex state ($g > -7.79$) and those of the corresponding
modes above the low-lying soliton state ($g < -7.79$).
(c) The number of virtually excited atoms in the Bogoliubov modes.
The inset is the logarithmic plot of the excitation energies and the
number of excited atoms (both of them refer to the common scale) as
functions of $|g - g_Q^{\rm cr}|$.
The vertical dotted lines in (a)-(c) show $g_Q^{\rm cr} = -7.79$.
}
\label{f:lowly}
\end{figure}

Using the Newton-Raphson method~\cite{Edwards} starting from appropriate 
initial states, we obtain the stationary states in the frame of reference
rotating at frequency $\Omega$.
Figure~\ref{f:lowly} shows various properties of the axisymmetric vortex 
state and the low-lying state.
When $|g|$ is increased, the stationary state bifurcates at the critical
point $g^{\rm cr}_Q = -7.79$ of the quadrupole instability, at which the
low-lying state emerges as represented by the lower branches of $E$ and
$\mu$ in Fig.~\ref{f:lowly} (a).
Density profiles of the low-lying states for $g = -8$ and $g = -10$
are shown as gray-scale images in Fig.~\ref{f:lowly} (a).
It is clear from them that the split-merge cycle is due to oscillations
between the axisymmetric vortex state (upper branch) and the low-lying
soliton state (lower branch).
It is interesting to note that the energies of the axisymmetric and
low-lying states are smoothly connected at the critical point, while the
chemical potential has a kink, which implies that the nature of the phase
transition between the vortex state and the soliton state is of second
order.

The excitation energies are plotted in Fig.~\ref{f:lowly} (b),
where $E_{\delta m}$ denotes the energy of the lowest excitation with
angular momentum $1 + \delta m$ above the axisymmetric vortex state.
As $g$ approaches $g^{\rm cr}_Q = -7.79$ in the axisymmetric regime, the
excitation energy $E_{\delta m = 2} + E_{\delta m = -2}$ vanishes, despite
the fact that both $E_{\delta m = 2}$ and $E_{\delta m = 2}$ are nonzero
at $g^{\rm cr}_Q$.
This implies that quasiparticles with $\delta m = \pm 2$ can be excited
in pairs without changing energy and angular momentum.
Hence, the two modes $\delta m = \pm 2$ may be strongly coupled via the
pair excitations.
If we follow the axisymmetric branch [the upper branch in
Fig.~\ref{f:lowly} (a)], the eigenenergies of the two modes become
complex.
If we follow the low-lying soliton branch, one of the two modes becomes
the Goldstone mode associated with the axisymmetry breaking, and the
energy of the other mode $E_1$ increases with decreasing $g$.

Figure~\ref{f:lowly} (c) shows the number of virtually excited atoms
$\int d{\bf r} |v|^2$ in the quadrupole modes $N_{\delta m = 2} +
N_{\delta m = -2}$ for $g > g^{\rm cr}_Q$ and that in the corresponding
mode $N_1$ for $g < g^{\rm cr}_Q$.
At $g^{\rm cr}_Q$ the number of excited atoms diverges, indicating that
the Bogoliubov approximation breaks down.
The inset of Fig.~\ref{f:lowly} (c) shows the logarithmic plot of the
excitation energies and the numbers of excited atoms.
Close to the critical point $|g - g^{\rm cr}_Q| \ll 1$,
the excitation energies are found to be proportional to $|g - g^{\rm
cr}_Q|^{1/2}$ and the numbers of excited atoms to $|g - g^{\rm
cr}_Q|^{-1/2}$.

We note that these behaviors near the critical point (the kink in $\mu$,
the softening of the excitation mode, the divergence in the number of
excited atoms, and their power laws) are very similar to those occurring
in the transition between the uniform-density and soliton states in a 1D
ring~\cite{Kanamoto}.
Thus, in the immediate vicinity of the critical point at which the vortex 
split occurs, the mean-field and Bogoliubov approximations break down due
to large quantum fluctuations, and we must employ, e.g., the exact
diagonalization method~\cite{Kanamoto} to study this regime.
However, the mean-field theory used in the present paper is still valid
except for this narrow critical region.

\section{Conclusions}
\label{s:conclusion}

We studied dynamical instabilities in singly- and doubly-quantized vortex
states with attractive interactions, and found a rich variety of dynamics
triggered by the dynamical instabilities.
The dynamics depend on the strength of interaction $g$, the trap geometry
$\lambda$, the initial perturbations, and the manner (adiabatic or sudden)
in which $g$ changes.

A singly-quantized vortex in a trap with $\lambda \gtrsim 0.34$
first exhibits the quadrupole instability as $g$ is adiabatically
decreased.
The vortex then splits into two clusters revolving around the center of
the trap.
In a spherical trap the split fragments immediately collapse for any value
of $g$ that gives rise to the split instability, whereas in a
pancake-shaped trap with $\lambda = 10$ they unite to recover the original
vortex and then the split-merge cycles repeat for $|g|$ slightly
exceeding $|g_Q^{\rm cr}|$.
In a cigar-shaped trap with $\lambda \lesssim 0.34$, the dipole
instability causes the atoms to concentrate in one cluster and then the
system collapses.

The doubly-quantized vortex state is always dynamically unstable against
disintegration of the vortex core for $g < 0$, and therefore the
disintegration is unavoidable if we adiabatically decrease $g$.
If $g$ is suddenly changed to some negative value with an appropriate
initial perturbation to the condensate, the octupole instability can
dominate the quadrupole instability, and the vortex splits into three
clusters.
In a pancake-shaped trap, these undergo split-merge cycles as in the
case of a singly-quantized vortex, but eventually one of them collapses
due to the quadrupole instability that always exists.
We have also demonstrated that the split phenomena can be observed using
the current experimental setup of the MIT group.

To understand the dynamical instability and the split-merge cycles, we
performed a variational analysis.
The emergence of a complex eigenvalue in some modes was reproduced by the
Gaussian trial functions.
We were able to reproduce the split-merge cycles by assuming only three
relevant modes with angular momenta $1$ and $1 \pm 2$, suggesting that
only a few modes are associated with the split-merge cycles.

In the immediate vicinity of the critical strength of interaction of the
dynamical instability, the transition from the axisymmetric vortex state
to the split state is accompanied by divergence in the number of virtually
excited atoms and by softening of the excitation mode according to the
Bogoliubov theory, which is very similar to the 1D case discussed in
Ref.~\cite{Kanamoto}.
The dynamical instabilities in vortex states of attractive BECs therefore
involve a fundamental generic issue of the decay of a many-particle
quantum system.

Vortex-split phenomena are attributed to the interplay between the
attractive interaction and the topological phase defect, that is, the
condensate is bound to break axisymmetry, since they cannot gather at
the phase defect.
The emergence of the dynamical instabilities and ensuing dynamics may be
regarded as pattern formation due to symmetry breaking.
More complicated and interesting pattern might be formed in the vortex
lattice, which will be published elsewhere.

Dynamics after the collapse were not studied because 3D simulation of the
collapse and explosion phenomena exceeds our present computational
ability.
The burst atoms ejected from each cluster interfere with each other and
exhibit some anisotropic pattern if the burst atomic cloud is coherent.
Therefore experiments on vortex split and fragmented collapse will provide
a test of whether or not the burst atoms are coherent.

\begin{acknowledgments}
We thank J. K. Chin for discussions.
This work was supported by the Special Coordination Funds for Promoting
Science and Technology and a Grant-in-Aid for Scientific Research (Grant
No. 15340129) by the Ministry of Education, Science, Sports, and Culture
of Japan, and by the Yamada Science Foundation.
\end{acknowledgments}


\begin{thebibliography}{}

\bibitem{Matthews}
M. R. Matthews, B. P. Anderson, P. C. Haljan, D. S. Hall, C. E. Wieman,
and E. A. Cornell, Phys. Rev. Lett. {\bf 83}, 2498 (1999).

\bibitem{Madison}
K. W. Madison, F. Chevy, W. Wohlleben, and J. Dalibard,
Phys. Rev. Lett. {\bf 84}, 806 (2000).

\bibitem{Abo}
J. R. Abo-Shaeer, C. Raman, J. M. Vogels, and W. Ketterle, Science {\bf
292}, 476 (2001).

\bibitem{Inouye}
S. Inouye, M. R. Andrews, J. Stenger, H. -J. Miesner, D. M. Stamper-Kurn,
and W. Ketterle, Nature {\bf 392}, 151 (1998).

\bibitem{Cornish}
S. L. Cornish, N. R. Claussen, J. L. Roberts, E. A. Cornell, and
C. E. Wieman, Phys. Rev. Lett. {\bf 85}, 1795 (2000).

\bibitem{Williams}
J. E. Williams and M. J. Holland, Nature {\bf 401}, 568 (1999).

\bibitem{Nakahara}
M. Nakahara, T. Isoshima, K. Machida, S. -I. Ogawa, and T. Ohmi, Physica
(Amsterdam) {\bf 284-288B}, 17 (2000);
T. Isoshima, M. Nakahara, T. Ohmi, and K. Machida, Phys. Rev. A {\bf 61},
063610 (2000).

\bibitem{Leanhardt}
A. E. Leanhardt, A. G\"orlitz, A. P. Chikkatur, D. Kielpinski, Y. Shin,
D. E. Pritchard, and W. Ketterle, Phys. Rev. Lett. {\bf 89}, 190403
(2002).

\bibitem{Recati}
A. Recati, F. Zambelli, and S. Stringari, Phys. Rev. Lett. {\bf 86}, 377
(2001).

\bibitem{Sinha}
S. Sinha and Y. Castin, Phys. Rev. Lett. {\bf 87}, 190402 (2001).

\bibitem{Garcia00}
J. J. Garc\'{\i}a-Ripoll and V. M. P\'erez-Garc\'{\i}a,
Phys. Rev. Lett. {\bf 84}, 4264 (2000).

\bibitem{SaitoL}
H. Saito and M. Ueda, Phys. Rev. Lett. {\bf 89}, 190402 (2002).

\bibitem{Garcia99}
J. J. Garc\'{\i}a-Ripoll and V. M. P\'erez-Garc\'{\i}a, Phys. Rev. A {\bf
60}, 4864 (1999).

\bibitem{Mottonen}
M. M\"ott\"onen, T. Mizushima, T. Isoshima, M. M. Salomaa, and K. Machida,
Phys. Rev. A {\bf 68}, 023611 (2003).

\bibitem{Rosen}
P. Rosenbusch, V. Bretin, and J. Dalibard, Phys. Rev. Lett. {\bf 89},
200403 (2002).

\bibitem{Garciabend}
J. J. Garc\'{\i}a-Ripoll and V. M. P\'erez-Garc\'{\i}a, Phys. Rev. A {\bf
63}, 041603 (2001); {\it ibid.} {\bf 64}, 053611 (2001).

\bibitem{Engels}
P. Engels, I. Coddington, P. C. Haljan, V. Schweikhard, and E. A. Cornell,
Phys. Rev. Lett. {\bf 90}, 170405 (2003).

\bibitem{Pu}
H. Pu, C. K. Law, J. H. Eberly, and N. P. Bigelow, Phys. Rev. A {\bf 59},
1533 (1999).

\bibitem{Kanamoto}
R. Kanamoto, H. Saito, and M. Ueda, Phys. Rev. A {\bf 67}, 013608 (2003).

\bibitem{Sackett97}
C. A. Sackett, C. C. Bradley, M. Welling, and R. G. Hulet, Appl. Phys. B
{\bf 65}, 433 (1997).

\bibitem{Ueda98}
M. Ueda and A. J. Leggett, Phys. Rev. Lett. {\bf 80}, 1576 (1998).

\bibitem{SaitoA01}
H. Saito and M. Ueda, Phys. Rev. A {\bf 63}, 043601 (2001).

\bibitem{Edwards}
M. Edwards, R. J. Dodd, C. W. Clark, and K. Burnett,
J. Res. Natl. Inst. Stand. Technol. {\bf 101}, 553 (1996).

\bibitem{Dalfovo}
F. Dalfovo and S. Stringari, Phys. Rev. A {\bf 53}, 2477 (1996).

\bibitem{Ruprecht}
P. A. Ruprecht, M. J. Holland, K. Burnett, and M. Edwards, Phys. Rev. A
{\bf 51}, 4704 (1995).

\bibitem{Donley}
E. A. Donley, N. R. Claussen, S. L. Cornish, J. L. Roberts, E. A. Cornell, 
and C. E. Wieman, Nature {\bf 412}, 295 (2001).

\bibitem{SaitoLA}
H. Saito and M. Ueda, Phys. Rev. Lett. {\bf 86}, 1406 (2001);
Phys. Rev. A {\bf 65}, 033624 (2002).

\bibitem{Chin}
J. K. Chin, J. M. Vogels, and W. Ketterle, Phys. Rev. Lett. {\bf 90},
160405 (2003).

\bibitem{Castin}
Y. Castin and R. Dum, Eur. Phys. J. D {\bf 7}, 399 (1999).

\bibitem{Rokhsar}
D. S. Rokhsar, Phys. Rev. Lett. {\bf 79}, 2164 (1997).

\bibitem{Isoshima}
T. Isoshima and K. Machida, J. Phys. Soc. Jpn. {\bf 66}, 3502 (1997).

\bibitem{Fetter98}
A. L. Fetter, J. Low. Temp. Phys. {\bf 113}, 189 (1998).

\bibitem{Nozieres}
P. Nozi\'eres and D. Pines, {\it The Theory of Quantum Liquids}
(Addison-Wesley, Reading, MA, 1990).

\bibitem{Jitkee}
J. K. Chin, private communication.

\bibitem{Fetter}
A. L. Fetter, Phys. Rev. A {\bf 53}, 4245 (1996).

\end{thebibliography}
\end{document}